\documentstyle[preprint,aps]{revtex}

\begin{document}

\title
{Specific heat of a Fermi system near ferromagnetic quantum phase transition}

\author
{I. Grosu, D. Bodea and M. Crisan}

\address
{Department of Theoretical Physics, University of Cluj, 3400 Cluj-Napoca, Romania}

\maketitle

\begin{abstract}
We calculate the specific heat for an interacting Fermi system near the ferromagnetic phase
transition using the Renormalization Group method. The temperature dependence of the specific
heat present for dimension $D=3$ a logarithmic dependence which shows that the fermionic 
excitations reaches a non - Fermi behavior. The result is in good agreement with the 
experimental data obtained recently for $Ni_x Pd_{1-x}$ alloys.
\end{abstract}

\newpage
\narrowtext

\section{Introduction}

The occurrence of the non - Fermi behavior in the heavy - fermion system has been associated with
the antiferromagnetic quantum phase transition (AQPT) \cite{von,stockert,schoder,lenke} and is 
one of the most important subject concerning the non - Fermi behavior, introduced for the 
explanation of the normal state of cuprate superconductors. Recently the non - Fermi behavior 
was experimentally discovered in the systems as $Th_{1-x} U_x Cu_2 Si_2$ \cite{lenke} and the 
$Ni_x Pd_{1-x}$ alloy \cite{nicklas}, if $x<x_c$, $x_c$ being a critical concentration. This 
behavior has been also considered as a quantum phase transition (QPT),
but the systems present a ferromagnetic quantum phase
transition (FQPT). In this paper we will calculate the heat capacity for a model which describes
the interaction between itinerant - electron and magnetic moments. The Renormalization - Group
method (RG) used in previous papers \cite{hertz,millis,zulicke} will be adopted to describe the
QPT and the free energy calculated by RG will be used to show the existence of the logarithmic
term ($T \ln T$) which represent the non - Fermi behavior of the fermionic excitations.

The paper is organized as follows. In Sec. II we present the model and give the dynamic 
susceptibility of the system. Sec. III contain the RG scaling equations and in Sec. IV we
calculate the heat capacity. The relevance of the results compared with the other models is
discussed in Sec. V.

\section{Model}

The transition to the ferromagnetism due to the magnetic moments has been discussed by Doniach
and Wohlfart \cite{doniach} for the palladium ion. The basic idea of the model is the 
polarization of the itinerant - electrons by the magnetic impurity and the susceptibility was
calculated as:
\begin{equation}
\chi({\bf q}, \omega) = 
\frac{ \chi_0({\bf q}, \omega)}{1 - \left[ I + \frac{2J^2 R'}{J R - \omega} \right] 
\chi_0({\bf q}, \omega)}
\label{1}
\end{equation}
where $\chi_0({\bf q}, \omega)$ is the susceptibility of electrons given by:
\begin{equation}
\chi_0({\bf q}, \omega) = \frac{1}{N} \sum_k 
\frac{n_{k \uparrow} - n_{k+q \downarrow}}{\epsilon_k - \epsilon_{k+q} +  \Delta - \omega}
\label{2}
\end{equation}
$\Delta$ is the gap introduced by the electron - electron interaction $I$, and the 
electron - impurity interaction $J$, and was calculated as:
\begin{equation}
\Delta = I R + 2 J R'
\label{3}
\end{equation}
where
\begin{equation}
R = \frac{1}{N} \sum_k (n_{k \downarrow} - n_{k \uparrow})
\label{4}
\end{equation}
and
\begin{equation}
R' = - C <S^z>
\label{5}
\end{equation}
and $C$ is the concentration of the magnetic impurities with the spin ${\bf S}$.

Using the expansion:
\begin{equation}
\frac{\chi_0({\bf q}, \omega)}{\chi(0,0)} = 1 - Aq^2 - B \left( \frac{\omega}{q} \right)^2
+ \Delta \left[ D_1 \frac{\omega}{q^2} + D_2 \omega + \cdots \right] + i C \frac{\omega}{q}
\label{6}
\end{equation}
$A$, $B$, $C$, $D_i$ ($i=1,2$) being constants we get for susceptibility expressed by 
Eq. (\ref{1}) the form:
\begin{equation}
\chi({\bf q}, \omega) \simeq 
\frac{1}{\delta + aq^2 - b \frac{\Delta \omega}{q^2} - i \frac{\omega}{\Gamma q}}
\label{7}
\end{equation}
where $\delta$ is the concentration dependent parameter, given by:
\begin{equation}
\delta = {\chi(0,0)}^{-1} - F
\label{8}
\end{equation}
\begin{equation}
F = I + 2 J \frac{R'}{R}
\label{9}
\end{equation}
For $\Delta = 0$ Eq. (\ref{7}) approximate the dynamic susceptibility of the excitations in the 
itinerant ferromagnet (see Ref. \cite{hertz,millis}).

\section{Scaling equations}

The fluctuations of the magnetization in the critical region of the ferromagnetic state are
described by the action:
\begin{equation}
S_{eff} = S_{eff}^{(2)} + S_{eff}^{(4)}
\label{10}
\end{equation}
where
\begin{equation}
S_{eff}^{(2)} = \frac{1}{2} \sum_q {\chi(q)}^{-1} |\phi(q)|^2
\label{11}
\end{equation}
\begin{equation}
S_{eff}^{(4)} = \frac{u}{4} \prod_{i=1}^{4} \sum_{q_i} |\phi(q_i)|^4 \delta 
\left( \sum_{i=1}^{4} q_i \right)
\label{12}
\end{equation}
where
\begin{equation}
\sum_q = V T \sum_n \int \frac{d^d {\bf q}}{(2 \pi)^d}
\label{13}
\end{equation}
and $q_i = ({\bf q}_i, \omega_{ni})$, $\omega_{ni} = 2 \pi n T$ being the bosonic frequency and
$u$ the interaction, which satisfies $u>0$. Next we consider the scaling properties of this 
action. When we take the scale transformation $q' = q l$ and $\omega' = \omega l^z$, the 
Gaussian term will be transformed as:
\begin{eqnarray}
S_{eff}^{(2)} = \frac{1}{2} \sum_{q'} \left[ \delta + {q'}^2 l^{-2} + 
\left( \frac{|\omega_n|}{q'} \right) \Gamma l^{1-z} \right. \nonumber \\
\left. + b \Delta |\omega_n| l^{z-2} \right] |\phi(q')|^2
\label{14}
\end{eqnarray}
The fourth - order term, given by Eq. (\ref{12}) is scaled as:
\begin{equation}
S_{eff}^{(4)} = l^{4-(d+z)} u \prod_{i=1}^{4} \sum_{{q'}_i} \delta 
\left( \sum_{i=1}^4 {q'}_i \right) |\phi({q'}_i)|^4
\label{15}
\end{equation}
Therefore, the scaling equations should be:
\begin{equation}
\frac{d T(l)}{d l} = z T(l)
\label{16}
\end{equation}
\begin{equation}
\frac{d \Delta(l)}{d l} = (4-z) \Delta(l)
\label{17}
\end{equation}
\begin{equation}
\frac{d \Gamma(l)}{d l} = (z-3) \Gamma(l)
\label{18}
\end{equation}
\begin{equation}
\frac{d \delta(l)}{d l} = 2 \delta(l) + 2 u(l) (n+2) f_1
\label{19}
\end{equation}
\begin{equation}
\frac{d u(l)}{d l} = [4-(d+z)] u(l) - (n+8) u^2(l) f_2
\label{20}
\end{equation}
The Eq.(\ref{20}) shows that the quadratic term is scaled to zero for $d+z>4$. On the other hand
the existence of the ferromagnetic state is conditioned by the existence of the gap $\Delta$ and
from Eq. (\ref{17}) we may conclude that $z<4$. However, from (\ref{18}) we see that for $z>3$
the quantum fluctuations are important and for $z=3$, $\Gamma$ is dangerous irrelevant. These 
considerations leads to the conclusion that for $d=3$ the relation $d+z>4$ and $u$ indeed scales
to zero.

If only the Gaussian point of action $S_{eff}^{(2)}$ is considered, the partition function is
\begin{displaymath}
Z = {\bf Tr} \exp (- \beta F) \equiv \int D \phi^* D \phi \exp (S)
\end{displaymath}
where the free energy $F$ is given by:
\begin{equation}
F = V \int_0^{\Lambda} \frac{d^d q}{(2 \pi)^d} \int_0^{\Gamma_k} \frac{d \omega}{\pi}
n_B(\omega) \tan^{-1} \frac{\frac{\omega}{\Gamma q}}
{\delta + a q^2 - b \frac{\Delta \omega}{q^2}}
\label{21}
\end{equation}
$n_B(\omega)$ being the Bose function.
Performing the same scaling for $q$ and $\omega$ we obtain the equation for $F$:
\begin{equation}
\frac{d F(l)}{d l} = (d+z) F(l) + f_3 (T(l),\Delta(l),\delta(l))
\label{22}
\end{equation}

Following the method developed in Ref. \cite{millis,zulicke} we will calculate the free energy
and the specific heat for the two regimes which exist when we stop the scaling. These regimes
are the quantum regime ($T(l) \ll 1$) and the classical regime ($T(l) \gg 1$) and we expect an
important enhancement of the behavior in the specific heat due to the quantum effects.

\section{Specific heat}

The specific heat $C_v  = - T \partial^2 F / \partial T^2$ will be calculated from the general
solution of the Eq. (\ref{22}).
\begin{equation}
F(T) = \int_0^l d x e^{-(d+z)x} f_3 \left(T e^{zx} \right)
\label{23}
\end{equation}
following the method used in \cite{hertz,millis}.

The effect of the fluctuations is important in the quantum domain ($T(l) \ll 1$) and classical
domain ($T(l) \gg 1$). These two domains have been studied using the flow equations 
(see Ref. \cite{millis}) and the basic point is to stop the renormalization procedure when the
system is driven from the critical region in the region where we can perform a standard 
perturbative calculation. We introduce $l=l^*$ which near the critical point satisfies
$T(l^*) \ll 1$ for quantum domain and $T(l^*) \gg 1$ for the classical domain. This domains are
separated by a matching value $l_m$ defined by:
\begin{equation}
T(l_m) = 1
\label{24}
\end{equation}
calculated from Eq. (16) as:
\begin{equation}
l_m = \frac{1}{z} \ln \frac{1}{T}
\label{25}
\end{equation}

In this approximation we calculate the free energy from Eq. (\ref{23}) as:
\begin{equation}
F(T) = \int_0^{1/z \ln 1/T} d x e^{-(d+z)x} f_3 \left( T e^{zx} \right)
+ \int_{1/z \ln 1/T}^{l^*} d x e^{-(d+z)x} f_3 \left( T e^{zx} \right)
\label{26}
\end{equation}
Following Ref. \cite{millis,zulicke} we approximate $f_3(T) - f_3(0) \simeq A T^2$ in the low
temperature domain and $f_3(T) \simeq B T$ in the high temperature domain. The general form of 
the free energy given by (\ref{25}) is:
\begin{equation}
F(T) = F_1 T^2 \ln \frac{1}{T} - F_2 T
\label{27}
\end{equation}
and the specific heat obtained from Eq. (\ref{27}) has the form:
\begin{equation}
C_v = \gamma_0 T + \gamma_1 T \ln T
\label{28}
\end{equation}
In this equation the first term represent the effect of critical fluctuations in the classical 
domain and the second term describes the effect given by the quantum fluctuations in the 
proximity of the quantum critical point, where the system becomes non - Fermi in agreement
also with the resistivity temperature dependence \cite{lenke,nicklas}.

\section{Discussions}

We will discuss the results obtained in this paper in connection with the other approaches as 
well as to the experimental results.

First we mention that recently the problem of QPT in a ferromagnet was extensively studied
\cite{belitz1,belitz2,belitz3,belitz4} for a pure as well as for disordered ferromagnet. The
main idea of these papers is the occurrence of an effective long - range interaction between the
order parameter fluctuations. In fact we mention that this conclusion is valid only at $T = 0$ and
the non - analyticity is done to the fact that in \cite{belitz1,belitz2,belitz3,belitz4} the 
authors integrated out all the fermionic degrees of freedom \cite{sachdev}. However, in this case
the coupling to the electron - hole continuum is not any more possible, and the existent 
damping term is difficult to be explained in pure ferromagnet, at least due to this effect.

An important point is that such a non - analytic susceptibility does not respect the conservation
law for the magnetization in an itinerant - electron ferromagnet.  The difficulties due to the
application of the RG method as well as the conjecture that for $T \ne 0$ the model becomes
massive by the simplest substitution $q^{d-1} \rightarrow (q+T)^{d-1}$ determined us to use the
Hertz - Millis (HM) \cite{hertz,millis,zulicke} approach in description of the ferromagnetic
quantum phase transition. 

The formal inconvenience which appears by taking into consideration the occurrence of many energy
scales seems to be normal, but also natural if we are crossing from quantum -  to classical 
regime.

The occurrence of the non - Fermi behavior in this model is also a good reason to use the HM 
version of RG for this problem. The only problem which remain open is the value of $z$ in this
model, but we expect that the value from the disordered ferromagnet to be more appropriate.

Finally, we mention that we obtained a good agreement with the experimental data 
\cite{lenke,nicklas} for the specific heat and we also need an accurate calculation for the 
magnetic susceptibility as a function of temperature. A preliminary calculation showed a strong
deviation from normal Pauli behavior, but also from the Curie - like behavior as is expected
from self - consistent fluctuation theory \cite{morija}. The occurrence of a $\ln T$ - term 
predicted by experimental results in AQPT is
due to the non - Fermi behavior. The accurate analysis of this problem for FQPT is in progress.

\section*{Acknowledgements}

We thank professor A. - M. Tremblay for relevant observations of the physical features of the 
problem.


\begin{references}

\bibitem{von} H. von L\"{o}nheysen, J. Phys. Cond Mat. {\bf 8}, 9689 (1996).
\bibitem{stockert} O. Stockert, H. von L\"{o}nheysen, A. Rosch, N. Pyka and M. Loewenhampt, Phys. 
                   Rev. Lett. {\bf 80}, 5625 (1998).
\bibitem{schoder} A. Schoder, G. Aeppli, E. Bucher, R. Ramazashvili and P. Coleman, Phys. Rev. 
                  Lett. {\bf 80}, 5623 (1998).
\bibitem{lenke} M. Lenkewitz, S. Corsepius, G-F. von Blanckenhagen and G. R. Stewart, Phys. Rev.
                B {\bf 55} 6409 (1997).
\bibitem{nicklas} M. Nicklas, M. Brando, G. Knebel, F. Mayr, W. Trinke and A. Ladl, Phys. Rev.
                  Lett. {\bf 82}, 4286 (1999).
\bibitem{hertz} J. A. Hertz, Phys. Rev. B {\bf 14}, 1165 (1976).
\bibitem{millis} A. J. Millis, Phys. Rev. B {\bf 48}, 7183 (1993).
\bibitem{zulicke} U. Z\"{u}licke and A. J. Millis, Phys. Rev. B {\bf 51}, 8956 (1995).
\bibitem{doniach} S. Doniach and E. P. Wohlfart, Proc. R. Soc. London {\bf 296}, 442 (1967).
\bibitem{morija} T. Moriya, Spin Fluctuations in Itinerant Electron Magnetism, Springer-Verlach 
                 Berlin 1985.
\bibitem{belitz1} D. Belitz and T. Kirkpatrick, Phys. Rev. B {\bf 44}, 955 (1991).
\bibitem{belitz2} T. Kirkpatrick and D. Belitz, Phys. Rev. {\bf 45}, 3187 (1992); Phys. Rev. B
                  {\bf 53}, 14364 (1996).
\bibitem{belitz3} T. Vojta, D. Belitz, T. Kirkpatrick and R. Narayanan, Z. Phys. B 
                  {\bf 103}, 451 (1997).
\bibitem{belitz4} T. Kirkpatrick and D. Belitz, Phys. Rev. B {\bf 62}, 952 (2000).
\bibitem{sachdev} S. Sachdev "Quantum Phase Transitions", Cambridge University Press, 
                  England, 1999.

\end{references}
\end{document}